\documentclass[12pt,a4paper]{article}

  \usepackage{a4wide}
  \usepackage{latexsym}
  \usepackage{epsf}
  \usepackage{amssymb}
  \usepackage{graphicx}
  \usepackage{amsmath, cite}
  \usepackage{amsmath,amssymb,amsthm}
  \usepackage{verbatim}
  \usepackage{hyperref}
  \usepackage{bbold}
  \usepackage{units}

\renewcommand{\det}[1]{\textrm{det}\,{#1}}
\renewcommand{\d}{\textrm{d}}
\newcommand{\e}{\textrm{e}}

\newcommand{\SL}{\mathop{\rm SL}}
\newcommand{\SO}{\mathop{\rm SO}}
\newcommand{\SU}{\mathop{\rm SU}}
\newcommand{\Tr}{\text{Tr}}
\newcommand{\be}{\begin{equation}}
\newcommand{\ee}{\end{equation}}
\newcommand{\sgn}{\text{sgn}\,}

\begin{document}

\begin{center}

\begin{flushright}
{\small UUITP-06/13
 \\
} \normalsize

{\small WITS-CTP-112
 \\
} \normalsize
\end{flushright}
\vspace{0.3 cm}

{\LARGE \bf{Power-law cosmologies in minimal and maximal gauged supergravity\\\vspace{0.2cm}}}

\vspace{1.1 cm} {\large J. Bl{\aa}b\"ack$^a$, A. Borghese$^b$ and S. S. Haque$^c$}\\

\vspace{0.8 cm}{$^{a}$Institutionen f{\"o}r fysik och astronomi,\\ Uppsala Universitet, Uppsala, Sweden}\\
\vspace{0.2 cm}{$^{b}$Centre for Theoretical Physics, University of Groningen,\\Nijenborgh 4 9747 AG Groningen, The Netherlands}\\
\vspace{0.2 cm}{$^{c}$NITheP, School of Physics, and Centre for Theoretical Physics,\\University of the Witwatersrand,\\Johannesburg, WITS 2050, South Africa } \footnote{{\ttfamily { johan.blaback @ physics.uu.se, a.borghese@rug.nl, shajid.haque@wits.ac.za}}}\\

\vspace{0.8cm}

{\bf Abstract}
\end{center}

\begin{quotation}
\noindent In this paper we search for accelerating power-law solutions and ekpyrotic solutions within minimal and maximal four dimensional supergravity theories. We focus on the {\it STU} model for $\mathcal{N}=1$ and on the new ${\rm{CSO}}(p,q,r)$ theories, which were recently obtained exploiting electromagnetic duality, for $\mathcal{N}=8$. In the minimal case we find some new ekpyrotic solutions, while in the maximal case we find some new generic power-law solutions. We do not find any new accelerating solutions for these models.
\end{quotation}

\newpage

\tableofcontents

\section{Introduction}
One of the most striking discoveries of modern physics is the observed acceleration of our universe \cite{Riess:1998cb, Perlmutter:1998np}. From an effective field theory point of view, there are two main constructions that can describe cosmic acceleration. The first is a positive cosmological constant, and the second is the presence of some exotic energy density component (usually a homogeneous, running scalar field \cite{Halliwell:1986ja, Copeland:1997et}), originally called quintessence \cite{Caldwell:1997ii}. A further natural step consists in realising these effective models in the context of a UV complete theory such as string theory. 
There are several paths that can be pursued in doing this. For instance, one can try to obtain these effective field theories from the compactification of a higher dimensional theory \cite{Townsend:2001ea, Townsend:2003qv, Townsend:2003fx, Bergshoeff:2003vb}. In any case, in view of the close relationship with string theory, it is worth trying to embed these effective models in supergravity theories \cite{Kallosh:2001gr,VanProeyen:2006mf, Kallosh:2002gf}.

In this paper we describe some running scalar field models in minimal and maximal supergravity. In particular, we search for models realising accelerating power-law cosmologies. To be more specific, we will always analyse supergravity theories in which the scalar sector dynamics is driven by a scalar potential of the form
\begin{equation} \label{Multiple exponential}
V (\vec \phi)=\sum_{i} \Lambda_i \ \e^{\vec \alpha_i \cdot \vec \phi} \, ,
\end{equation}
where $\Lambda_i, \alpha_i$ are real numbers. We will always assume the space-time metric to be the FLRW one with scale factor $a(t),$ where $t$ is the cosmic time. A power-law cosmology will have $a(t) \propto t^{P}$. In some occasions, we will be able to write the scalar potential as
\begin{equation} \label{Single exponential}
V = e^{c \, \psi} \, U \, ,
\end{equation}
where $U$ is a function of the remaining scalar fields, which need to be stabilised. In such occasions, 
we might get power-law solutions, also known as scaling solutions. The coefficient  $c$ of the scalar $\psi$ in the above expression is related to the power-law behavior ($P=\nicefrac{1}{c^2}$). 
The scaling solution will be accelerating when $P > 1$ and $U$ is stabilised at a positive value \cite{Collinucci:2004iw, Hartong:2006rt}. As we explain in the main text, the search for accelerating scaling solutions has many similarities with the search for dS solutions. This can be understood by noting that torus reduction of a higher dimensional de Sitter solution gives an accelerating power-law solution \cite{Rosseel:2006fs}.

In the last few years there has been some extensive research done on the construction of de Sitter solutions in supergravity. This task can be accomplished more easily in the context of $\mathcal{N} = 1$ supergravity due to the larger freedom in building the theory. Examples are dS solutions in $\mathcal{N}=1$ supergravities that originate \cite{Silverstein:2007ac, Haque:2008jz,Caviezel:2008tf, Flauger:2008ad, Danielsson:2009ff, Danielsson:2010bc, Danielsson:2011au} from string theory at tree level. 
$\mathcal{N} = 2$ supergravity seems to have meta-stable de Sitter vacua \cite{Fre:2002pd, Catino:2013syn, Cosemans:2005sj}, but unfortunately the higher dimensional origin for those examples is not clear. In $\mathcal{N} = 4,8$, up to date, there is no known meta-stable dS critical point. All known dS critical points \cite{deRoo:2002jf, deRoo:2003rm, deRoo:2006ms, Gibbons:2001wy, Cvetic:2004km} develop some instability in the scalar spectrum. Furthermore, the tachyonic directions have usually a mass squared whose absolute value is proportional to the value of the potential at the critical point ($|m^{2}_{\textsc{tach}} | \propto V |_\star$). This is known as the $\eta$ problem for supergravity. 

In this paper we have considered the two extreme cases -- the minimal and the maximal supergravities. For arbitrary $\mathcal{N}=1$ theories, we can always get stable dS vacua and stable power-law solutions. However, we would like to analyse theories with a clear string theory origin in order to understand what kind of constraints we get on IR Lagrangians from UV completions. This is why we study the {\it STU} model. For such a model a complete dictionary between terms in the scalar potential and fluxes in type II string theory has been worked out. Exploiting string dualities it is then possible to extend the amount of fluxes one can turn on in the supergravity picture. Some of them, called geometric, have a clear higher dimensional origin, while others called non-geometric do not. Remarkably, stable dS vacua seem impossible to get unless one adds non geometric fluxes \cite{Blaback:2013ht}. 
In our paper, we have investigated whether the same obstacles occur for accelerating power-law solutions, and our results indicate that this is indeed the case. At the other end of the spectrum, we study $\mathcal{N} = 8$ supergravities. Even though these are not realistic theories from a particle physics point of view, they have some attractive features. In particular, the great amount of supersymmetry imposes the presence of a single supermultiplet i.e., the multiplet containing the graviton. Furthermore, for some special gaugings, the higher dimensional origin is clear. Again 
it looks like accelerating scaling solutions are as inconspicuous as de Sitter solutions. All these findings further add to the idea that string theory does not seem to favour an accelerated expansion.

Another important avenue of string theory research is how it relates to the cosmological inflation. This cosmological paradigm suggests that for a very short period of time, the early universe went through an enormous amount of acceleration. Within a fraction of second, it grew from a subatomic scale to a macroscopic scale. The most widely discussed alternative to the theory of cosmic inflation is the ekpyrotic or cyclic cosmology \cite{Steinhardt:2001st, Khoury:2001bz, Khoury:2003rt}. It proposes that the current expansion of our universe is one out of an infinite number of cycles. Each cycle begins with
a big bang phase, and then a slowly accelerating expansion phase, followed by a slow contraction phase, finally ending with a big crunch. In this paper, we have also searched for these ekpyrotic solutions within maximal and minimal supergravity theories. Searching for ekpyrotic solutions is mathematically very similar to the search for accelerating scaling solutions. The same exponential potential (\ref{Single exponential}) could be used for ekpyrosis if it is very steep ($P < \nicefrac{1}{3}$) and $U$ is stabilised at a negative value \cite{Khoury:2004xi} in order to avoid growing anisotropies at the singularity \cite{Creminelli:2006xe, Erickson:2003zm}.

As anticipated, for $\mathcal{N}=1,$ we will study the so called {\it STU} supergravity model \cite{Kachru:2002he, Derendinger:2004jn, DeWolfe:2004ns, Camara:2005dc, Villadoro:2005cu, Derendinger:2005ph, DeWolfe:2005uu, Aldazabal:2006up, Aldazabal:2007sn}. In this minimal supergravity model the effective superpotential is generated from fluxes \cite {Giddings:2001yu}. In\cite{Dibitetto:2011gm} the complete vacuum structure of the SO(3) truncation of $\mathcal {N} = 4$ supergravity was analysed.  These half-maximal supergravity theories arise as the low energy limit of certain type IIA orientifold compactifications with background fluxes, D6-branes and O6-planes.
Using the same technique for the $\mathcal{N}=4$ theory the authors of \cite{Dibitetto:2011gm} worked out the complete set of solutions of type IIA geometric backgrounds compatible with minimal supersymmetry. They showed that the SO(3) truncation also admits an $\mathcal{N} = 1$ superpotential formulation. The quadratic constraints coming from the consistency of the $\mathcal{N} = 4$ gauging give rise to three constraints on the fluxes. Whenever one relaxes one of these constraints, the model admits an $\mathcal{N}=1$ description. For this minimal supergravity model we have done an exhaustive search for power-law solutions. We have found three stable ekpyrotic solutions, however, have not found any accelerating scaling solution.

A systematic search for scaling solutions in gauged maximal supergravity was carried out in \cite{Rosseel:2006fs}. The search was conducted within a specific class of gaugings, namely the ${\rm{CSO}}(p,q,r)$ ones. The authors were able to find two scaling solutions with $P = 3  \ \text{and}\ 7$. The first one was already found in \cite{Townsend:2001ea} in the ${\rm{CSO}}(3,3,2)$ gauged theory and its higher dimensional origin was shown. The second one was in the ${\rm{CSO}}(4,3,1)$ gauged theory. In general, the ${\rm{CSO}}(p,q,r)$ gauged maximal supergravities can be obtained by non-compactification of 11 dimensional supergravity on hyperbolic internal manifolds \cite{Hull:1988jw}. 

The results of \cite{Rosseel:2006fs} were quite conclusive up to last year, at least for what concerns the ${\rm{CSO}}(p,q,r)$ gaugings. Surprisingly, in \cite{Dall'Agata:2012bb} it has been discovered that there is not just a single maximal supergravity for every ${\rm{CSO}}(p,q,r),$ but rather a one parameter family of inequivalent theories with different physical features such as the number of critical points. From now on, we will denote the parameter with a phase $\omega$ such that, for $\omega = 0,$ we recover the old supergravities. In particular, in \cite{Dall'Agata:2012sx} it has been discovered that in the ${\rm{SO}}(4,4)$ gauged theory dS critical points in theories with $\omega \neq 0$ have tachyons whose mass can be made as small as one wants, thus solving the $\eta$ problem. In other words, even though the critical points remain unstable, the instability is milder. 

Motivated by these findings, we sought to explore these newly constructed theories for scaling solutions. Despite the expectations, we found out that there are no new accelerating or ekpyrotic solutions for these new theories but only some generic power-law solutions.

\section {Minimal Supergravity ($\mathcal{N}=1$)}

The {\it STU} supergravity model \cite{Kachru:2002he, Derendinger:2004jn, DeWolfe:2004ns, Camara:2005dc, Villadoro:2005cu, Derendinger:2005ph, DeWolfe:2005uu, Aldazabal:2006up, Aldazabal:2007sn} originates from the SO(3) truncation of the half maximal supergravity in four dimensions. They consist of  three complex scalar fields $S, T, \ \text{and} \ U$.  This half maximal supergravity ($\mathcal{N}=4$) arises as a low energy limit of massive type IIA orientifold compactifications on a $T^6/(\mathbb{Z}_2\times \mathbb{Z}_2)$ orbifold. In \cite{Dibitetto:2011gm}, the authors showed that the SO(3) truncation also admits an $\mathcal{N} = 1$ superpotential formulation in terms of a real K{\"a}hler potential  $K$ and a holomorphic superpotential $W.$ In this section, we will explore this minimal supergravity model searching for cosmological power-law solutions.  

The scalar potential \label{potential} can be written from the K{\"a}hler potential and the superpotential as
\be \label{from sp}
V= e^K \left[  \sum_\Phi K ^{\Phi \bar\Phi} |D_{\Phi} W|^2- 3 | W |^2 \right],
\ee
$K ^{\Phi \bar\Phi} $ is the inverse of the K{\"a}hler metric $K _{I\bar J} = \frac{\partial K}{\partial \Phi^I \partial \bar\Phi^{\bar J}}$ and $D_{\Phi} W= \frac{\partial W}{\partial \Phi}+\frac{\partial K} {\partial \Phi} W $. In terms of complex scalars $S,T,U,$ the K{\"a}hler potential and superpotential are
\begin{align}
&K=-\log\{-i (S-\bar S)\}-3 \log\{-i(T-\bar T)\}-3 \log\{-i(U-\bar U)\}\,,\nonumber\\
& W=a_0-3 a_1 U+3 a_2 U^2-a_3U^3-b_0S+3 b_1 S U +3 c_0 T+3(2 c_1-\tilde c_1) T U\,.
\end{align}
We keep the fluxes that are part of the geometric type IIA duality frame. The origins of these fluxes are listed in Table 1.
\begin{table}[!ht] 
\centering \begin{tabular}{c c} 
\hline\hline 
Type & $\phantom{\dfrac{1}{2}}$Flux no.$\phantom{\dfrac{1}{2}}$ \\[1mm] 
\hline \\[-8mm] & \\ 
NSNS &$a_0,a_1,   a_2,   a_3$ \\[1mm]
RR&$b_0,c_0$ \\[1mm]
Metric&$c_1,b_1,\tilde c_1$ \\[1mm] 
\hline\hline \end{tabular} 
\caption{Flux origins} 
\end{table} \\
These fluxes are not entirely independent of each other -- there are quadratic constraints 
that give rise to the following flux relations \cite{Dibitetto:2011gm}
\begin{align}
& c_1(c_1-\tilde c_1)=0\,,\nonumber\\
& b_1 (c_1-\tilde c_1)=0 \,.
\end{align}
When we search for different types of solutions, we will solve these flux conditions by $c_1 = \tilde c_1$, which is not a restriction, since it turns out that the superpotential only depends on $(2c_1 - \tilde c_1)$.

The complex scalars $S, T, U$ are constructed from the real scalars as follows
\be S= \chi+i e^{-\phi},  \   T= \chi_1+ i e^{-\varphi_1}, \  U=\chi_2+ i e^{-\varphi_2}.\ee
The kinetic term of the Lagrangian can be written in terms of the complex fields as
\be
\mathcal{L}_{kin}=K_{I \bar J} \partial \Phi \partial \bar \Phi=\frac{\partial S \partial \bar S}{\{-i(S-\bar S)\}^2}+3 \frac{\partial T \partial \bar T}{\{-i(T-\bar T)\}^2}+3 \frac{\partial U \partial \bar U}{\{-i(U-\bar U)\}^2}.
\ee
We perform the following normalization in order to get the dilatonic fields in canonically normalized form 
\be
 \phi \to \sqrt{2}\ \phi_1\,, \quad  \varphi_1 \to \sqrt{\nicefrac{2}{3}}\ \phi_2\,, \quad \varphi_2 \to \sqrt{\nicefrac{2}{3}}\ \phi_3\,.
\ee

In our search for power-law solutions, we start by rewriting the full potential (\ref{from sp}) in terms of these three real canonically normalised dilatons ($\phi$'s) and three real axions ($\chi$'s). 
It takes the following form
\be
V (\vec \phi,\chi_j)=\sum_{i=1}^{11} \Lambda_i (\chi_j)  \ \e^{\vec \alpha_i. \vec \phi},
\ee
where $\vec\phi=\{\phi_1,\phi_2,\phi_3\}$ and $\chi_j=(\chi,\chi_1,\chi_2)$ and 
the scalar potential is characterised by 11 vectors $\vec \alpha$ listed in the Appendix (\ref{app:stu}). After that, we need to factor out a scalar field and express the potential in the following form

\be
V= e^{c \psi_1}U(\psi_2, \psi_3, \chi_j),
\ee
we recall that the factor $c$ corresponds to the power-law behaviour, $P=\frac{1}{c^2}$ with the scale factor $a(t)\sim t^P$. For accelerating power-law solutions, we want $P>1$, and we need to stabilize the potential 
 to a positive minimum, whereas ekpyrotic solutions require a negative potential with $P<\frac{1}{3}$. We will refer to any other solution with a positive minimum and  $1<P<\frac{1}{3}$ as a generic scaling solution. In the following section, we will search for these solutions. 
\subsection{Approach}

In order to analyse all the possible ways of constructing a scaling solution, we need to study all possible combinations of $\vec{\alpha}_i$ \footnote{The list of $\alpha$ vectors for the full scalar potential is in the Appendix (\ref{app:stu})} 
that are mutually affine. The exact definition of affine $\vec{\alpha}_i$ sets can be found in \cite{Hartong:2006rt}. Essentially, it means that after a field rotation, it is possible to extract an overall exponential factor for at least one field. Furthermore, we only need to study the largest common sets of $\alpha$ vectors that are mutually affine. For example, if ${\vec{\alpha}_1,\ldots,\vec{\alpha}_4}$ and ${\vec{\alpha}_1,\ldots,\vec{\alpha}_5}$ are both sets of mutually affine vectors, then we only need to study the latter.

These largest possible sets of mutually affine vectors end up being $17$. The $P$ for the sets are listed below.
\begin{equation}
\left\{ \frac{3}{2},\frac{7}{2},\frac{7}{2} \big| \frac{1}{6},\frac{2}{9},\frac{13}{54},\frac{17}{64},\frac{5}{16},\frac{31}{98}\big|\frac{1}{3}, \frac{19}{50},\frac{7}{18},\frac{7}{18},\frac{1}{2},\frac{1}{2},\frac{1}{2},\frac{5}{6}\right\}.
\end{equation}
To only keep a specific set of mutually affine vectors, we need to eliminate other terms, which can be done by choosing the appropriate values for the fluxes. If the fluxes chosen also make additional terms vanish, in general the $P$-value is lowered. This means that only three of the above are possible candidates for accelerating scaling solutions with $P>1$, but every single set of mutually affine vectors is a candidate for ekpyrotic solutions. 

The list above for possible $P$ values has three segments. The first segment with 3 cases has $P>1,$ which can provide us with accelerating solutions. The second segment has 6 cases and corresponds to $P<\frac{1}{3}$ which we will consider for the ekpyrotic solutions. Finally the last segment has 8 cases and corresponds to $1>P\ge \frac{1}{3}$ that we will consider for generic solutions. We will now proceed to analyse the above cases to see if they allow for accelerated, ekpyrotic or a generic scaling solution.

\subsection{Candidates for acceleration: Positive potential with $P>1$}

\begin{table}[ht] 
\centering \begin{tabular}{c c c} 
\hline\hline 
Case no.&Affine set of $\alpha$-vectors & $\phantom{\dfrac{1}{2}}P$\\[1mm] 
\hline \\[-8mm] & & \\
1&$\{\vec \alpha_3,\vec\alpha_4,\vec \alpha_5,\vec \alpha_{10}\}$&$\nicefrac{3}{2}$ \\[1mm]
2&$\{\vec \alpha_2,\vec\alpha_4,\vec \alpha_8,\vec \alpha_{10}\}$&$\nicefrac{7}{2}$ \\[1mm] 
3&$\{\vec \alpha_3,\vec\alpha_4,\vec \alpha_5,\vec \alpha_{11}\}$&$\nicefrac{7}{2}$ \\[1mm] 
\hline\hline \end{tabular} 
\caption{Accelerated cases} 
\end{table}
We now begin to analyse the candidates for accelerating solution. The analysis is performed in the following manner. First, we determine the conditions that we are forced to maintain if we want to keep a desired set of vectors. Then, the axions are stabilized. After that, we control the remaining exponential terms and analyse what solution it can support. 

Now from the three cases listed in Table 2, the only possible solution that we can construct from the $\alpha$ vectors is with the subset ${\vec{\alpha}_3}$. This is, however, an ekpyrotic solution since the single ${\vec{\alpha}_3}$ corresponds to a $P=\frac{3}{10}$.\\

{\bf Solution corresponding to the set $\{\vec{\alpha}_3\}$ with $P=\frac{3}{10}:$} The fluxes are 
\begin{equation}
a_1 = \frac{c_1 a_0}{3\ c_0},\quad a_2 = 0,\quad a_3 = 0, \quad b_0 = 0,\quad b_1 = 0,
\end{equation}
and the axions have extremal points at
\begin{equation}
\chi^\star \ \  \textrm{decouples}, \quad \chi^\star_1 =  - \frac{a_0}{3 c_0}, \quad \chi^\star_2 = \frac{c_0}{{c}_1}.
\end{equation}
These potentials have the following form at the extremal point for the axions
\begin{equation}
\left.V\right|_{\star} = -\frac{3 c_1^2}{32} \  e^{\sqrt{2} {\phi_1}+\sqrt{\frac{2}{3}} ({\phi_2}+{\phi_3})},
\end{equation}
which has $P=\frac{3}{10}$ and the potential is negative. It is a scaling solution in the direction of the field
\begin{equation}
\psi_1 = \sqrt{\frac{3}{5}} \phi_1 + \sqrt{\frac{1}{5}} (\phi_2+\phi_3). 
\end{equation}
This solution is possible due to the presence of $\vec{\alpha}_3$, and any set of affine vectors that contain this $\alpha$-vector should display this ekpyrotic solution. \\


\subsection{Candidates for ekpyrosis: Negative potential with $P<\frac{1}{3}$} 

\begin{table}[ht] 
\centering \begin{tabular}{c c c} 
\hline\hline 
Case no.& Affine set of $\alpha$-vectors& $\phantom{\dfrac{1}{2}}P$ \\[1mm] 
\hline \\[-8mm] & & \\
1&$\{\vec \alpha_2,\vec\alpha_4,\vec \alpha_6,\vec \alpha_{9}\}$&$\nicefrac{31}{98}$ \\[1mm]
2&$\{\vec \alpha_1,\vec\alpha_5,\vec \alpha_7,\vec \alpha_{9}\}$&$\nicefrac{2}{9}$ \\[1mm] 
3&$\{\vec \alpha_2,\vec\alpha_3,\vec \alpha_7,\vec \alpha_{8}\}$&$\nicefrac{5}{16}$ \\[1mm] 
4&$\{\vec \alpha_2,\vec\alpha_5,\vec \alpha_6,\vec \alpha_{9}\}$&$\nicefrac{17}{64}$ \\[1mm] 
5&$\{\vec \alpha_1,\vec\alpha_2,\vec \alpha_6,\vec \alpha_{8},\vec \alpha_{9}\}$&$\nicefrac{13}{54}$ \\[1mm]
6&$\{\vec \alpha_1,\vec\alpha_5,\vec \alpha_6,\vec \alpha_{7},\vec \alpha_{8},\vec \alpha_{10},\vec\alpha_{11}\}$&$\nicefrac{1}{6}$ \\[1mm]
\hline\hline \end{tabular} 
\caption{Ekpyrotic cases} 
\end{table}
Analysing the cases listed in Table 3, we were able to produce three different ekpyrotic solutions. One of them corresponds to the set of vectors with the subset $\{\vec{\alpha}_3\}$ that we have already found. The other two new ekpyrotic solutions are listed in the sections below.\\

{\bf Solution corresponding to the set $\{ \vec{\alpha}_2,\vec{\alpha}_6,\vec{\alpha}_9\}$ with $P=\frac{5}{22}:$} It is possible to construct ekpyrotic solutions for this set of affine vectors. One must choose the following values for the fluxes:
\begin{equation}
a_1 a_3 = a_2^2, \quad b_0 = 0, \quad b_1 = 0,\quad c_1 = 0.
\end{equation}
It is then possible to find extremal points 
for the axions at
\begin{equation}
\chi^\star\ \  \textrm{decouples}, \quad \chi^\star_1 = \frac{1}{3 c_0} \left( \frac{a_1 a_2}{a_3} - a_0 \right) , \quad \chi^\star_2 = \frac{a_2}{a_3}.
\end{equation}
The scaling then occurs in the direction of
\begin{equation}
\psi_1 = \frac{1}{\sqrt{55}} \left( 5 \phi_1 + 3 \sqrt{3} \phi_2 + \sqrt{3} \phi_3\right)
\end{equation}
with $P=\frac{5}{22}$.
The remaining dilatonic directions are
\begin{equation}
\begin{split}
\psi_2 &= \frac{1}{\sqrt{10}} \left(\phi_2 - 3 \phi_3\right)\\
\psi_3 &= \frac{1}{\sqrt{22}} \left (-2 \sqrt{3} \phi_1 + 3 \phi_2 + \phi_3\right),
\end{split}
\end{equation}
where $\psi_3$ decouples and $\psi_2$ have an extrema at
\begin{equation}
e^{2 \sqrt{\frac{5}{3}}\psi_2^\star} = \frac{2 c_0}{a_3}.
\end{equation}
The potential at the extremal point looks like
\begin{equation}
\left.V\right|_\star = -\left(\frac{5 a_3 c_0}{32\ 2^{\nicefrac{4}{5}}} \right )\left( \frac{c_0}{a_3}\right)^{\frac{1}{5}} e^{\sqrt{2} \phi_1 + \frac{3\sqrt{6}}{5} \phi_2 + \frac{\sqrt{6}}{5} \phi_3}.
\end{equation}
This solution is possible because of the set $\{ \vec{\alpha}_2,\vec{\alpha}_6,\vec{\alpha}_9\}$. Hence, any case that has these $\vec \alpha$ vectors as a subset should display the above ekpyrotic solution.\\

{\bf Solution corresponding to the set $\{ \vec\alpha_1,\vec{\alpha}_2,\vec{\alpha}_6,\vec\alpha_8,\vec{\alpha}_9\}$ with $P=\frac{13}{54}:$}
This contains the set $\{ \vec{\alpha}_2,\vec{\alpha}_6,\vec{\alpha}_9\}$, and provides a generalization to that ekpyrotic solution. The fluxes are now
\begin{equation}
a_1 a_3 = a_2^2, \quad b_1 = 0,\quad c_1 = 0,
\end{equation}
i.e. the same except $b_0$ remains. This makes $\chi_1$ not decouple, that is,
\begin{equation}
b_0 \chi^\star - 3 c_0 \chi^\star_1 =  a_0 - \frac{a_1 a_2}{a_3} , \quad \chi^\star_2 = \frac{a_2}{a_3}. 
\end{equation}
All the special cases, where any of $\{a_3, b_0, c_0\}$ are zero, either give positive potential or reproduce already studied cases. This means that below we only study the full set of affine vectors. We make the rotation into the fields
\begin{equation}
\begin{split}
\psi_1 &= \frac{1}{\sqrt{13}} \left( \sqrt{3} \phi_1 + 3 \phi_2 + \phi_3 \right),\\
\psi_2 &= \frac{1}{2}\left( - \sqrt{3} \phi_1 + \phi_2\right),\\
\psi_3 &= -\frac{1}{2 \sqrt{13}}\left( \phi_1 + \sqrt{3} \phi_2 - 4 \sqrt{3} \phi_3\right).
\end{split}
\end{equation} 
The scaling is along $\psi_1$ and $P=\frac{13}{54}$. It is possible to find extremal values for both remaining dilatons


\begin{equation}
e^{ 2 \sqrt{\frac{2}{3} }\ \psi_2^\star} = - \frac{4c_0}{b_0}, \quad e^{ 2 \sqrt{26} \ \psi_3^\star} = - \frac{4 a_3^4}{625 \ b_0 c_0^3} 
\end{equation}
and the potential is
\begin{equation}
\left. V \right|_\star = - \left (\frac{13 a_3^2}{16\ 2^{\nicefrac{8}{13}}\ 5^{\nicefrac{10}{13}}} \right) \left( -\frac{b_0 c_0^3}{a_3^4} \right)^{\frac{4}{13}} \ \e^{(\nicefrac{3}{13})(3\sqrt{2}\phi_1 + 3 \sqrt{6} \phi_2+ \sqrt{6}\phi_3)}.
\end{equation}
To have everything real and to get a negative potential, we must choose the following signs for fluxes: \  $\sgn c_0 = - \sgn b_0 = \sgn a_3 $. 

\subsection{Candidates for generic scaling solutions: Positive potential with $\frac{1}{3} < P < 1 $} \ \
\begin{table}[ht] 

\centering \begin{tabular}{c c c} 
\hline\hline 
Case no.&Affine set of $\alpha$-vectors& $\phantom{\dfrac{1}{2}}P$\\[1mm] 
\hline \\[-8mm] & & \\
1&$\{\vec \alpha_1,\vec\alpha_4,\vec \alpha_6,\vec \alpha_{8}\}$&$\nicefrac{5}{6}$ \\[1mm]
2&$\{\vec \alpha_1,\vec\alpha_4,\vec \alpha_9,\vec \alpha_{10}\}$&$\nicefrac{7}{18}$ \\[1mm]
3&$\{\vec \alpha_3,\vec\alpha_4,\vec \alpha_5,\vec \alpha_{6}\}$&$\nicefrac{19}{50}$ \\[1mm]
4&$\{\vec \alpha_1,\vec\alpha_3,\vec \alpha_6,\vec \alpha_{8}\}$&$\nicefrac{1}{3}$ \\[1mm]
5&$\{\vec \alpha_3,\vec\alpha_4,\vec \alpha_5,\vec \alpha_{8},\vec \alpha_{9}\}$&$\nicefrac{1}{2}$ \\[1mm]
6&$\{\vec \alpha_4,\vec\alpha_6,\vec \alpha_7,\vec \alpha_{10},\vec \alpha_{11}\}$&$\nicefrac{1}{2}$ \\[1mm]
7&$\{\vec \alpha_1,\vec\alpha_2,\vec \alpha_3,\vec \alpha_{4},\vec \alpha_{5},\vec \alpha_{7}\}$&$\nicefrac{7}{18}$ \\[1mm] 
8&$\{\vec \alpha_2,\vec\alpha_3,\vec \alpha_6,\vec \alpha_{7},\vec \alpha_{9},\vec \alpha_{10},\vec\alpha_{11}\}$&$\nicefrac{1}{2}$ \\[1mm]
\hline \hline \end{tabular} 
\caption{Generic cases} 

\end{table}

Finally, we explore the possible generic scaling solution candidates. The analysis is performed in the similar fashion as the accelerated case. There are eight different cases (listed in Table 4) but none of them gives a generic scaling solution. They only display the ekpyrotic solutions already found in the previous section. 


\section{Maximal Supergravity ($\mathcal{N}=8)$} 

In this section, we will first review the old ${\rm{CSO}}(p,q,r)$ gauged supergravity and the scaling solutions found therein following \cite{Rosseel:2006fs}. This will allow us to explain the main idea and set some conventions. Then we will go to the new maximal supergravities \cite{Dall'Agata:2012bb, Inverso:2012hs, Dall'Agata:2012sx, Borghese:2012zs} and classify the different theories with an eye on the possibility of finding new scaling or ekpyrotic solutions.

\subsection{The power-law solutions for the old theories}
We consider the truncation of the coset $E_{7(+7)}/\SU(8)$ to $\SL(8)/\SO(8)$ described by the symmetric matrix $\mathcal{M}$ of unit determinant, $\det{\mathcal{M}}=1$. In the truncation, the number of scalar fields is halved. The 35 scalars we are left with are divided in 7 dilatons (related to the Cartan generators of $\SL(8)$) and 28 axions. It is possible to further truncate the theory to the dilaton sector by virtue of the fact that the axions enter the equations of motion only quadratically. \\ 
Whenever a subgroup ${\rm{CSO}}(p,q,r) \subset \SL(8)$ is gauged, a scalar potential is generated of the form
\begin{equation}
V_{\textsc{old}}[\eta, \mathcal{M}] =  \Tr \big\{ \big( \eta \, \mathcal{M} \big)^2 \big\} -\tfrac{1}{2}\big( \Tr \big\{ \eta \, \mathcal{M} \big\} \big)^2 \, ,
\end{equation}
where \label {eta def}
\be
\eta=\begin{pmatrix} \mathbb{1}_p & & \\
& -  \mathbb{1}_q &\\
& &  \mathbb{0}_r
\end{pmatrix}\,.
\ee
A further consistent truncation in which we keep only $\SO(p)\times \SO(q)\times\SO(r)$ singlets is given by a two-scalar system, defined by $\varphi_1, \varphi_2$:
\be
\mathcal{M} = \e^{r \, \varphi_1} \begin{pmatrix} \e^{q \, \varphi_2} \, \mathbb{1}_p & & \\
&  \e^{-p \, \varphi_2} \, \mathbb{1}_q &\\
& &  \e^{-8 \, \varphi_1} \, \mathbb{1}_r
\end{pmatrix} \,.
\ee
The potential then becomes
\begin{equation}\label{standard1}
V_{\textsc{old}}(\varphi_1, \varphi_2) = \e^{2 \, r \, \varphi_1} \, U_{(p,q)}(\varphi_2) \, ,
\end{equation}
with 
\be \label{standard2}
U_{(p,q)}(\varphi_2) =  p \, (1-\tfrac{1}{2} \, p) \, \e^{2\, q \, \varphi_2} + q \, (1-\tfrac{1}{2} \, q) \, \e^{-2 \, p \, \varphi_2} + p \, q \, \e^{(q-p) \, \varphi_2} \, .
\ee
We are in front of a scaling solution if the following requests are satisfied:
\begin{enumerate}
\item There exists a \emph{positive} critical point of $U_{(p,q)}(\varphi_2)$: $ \quad \partial_{\varphi_2} U_{(p,q)} = 0 \, ,\quad U_{(p,q)} > 0$.
\item In terms of a canonical scalar $\phi_1$, we want the overall exponential factor $\e^{c \, \phi_1}$ to obey $c^2 <3$. The scale factor is then given by  $a(t)= t^{\frac{1}{c^2}}$. 
\end{enumerate}
For future purposes, let us study the features of $U_{(a,b)}$ for different values of $p$ and $q$. We find the following results (of course, the cases $(a, b)$ are equivalent to $(b,a)$ so we do not write it twice):
\begin{align}
\begin{array}{ccc}
U_{(a,b)} & \text{constant and negative} & (a, b) = \{(7, 0), (6, 0), (5, 0), (4, 0) , (3, 0) \} \, , \\[2mm]
U_{(a,b)} & \text{constant and equal to zero} & (a, b) =\{ (2, 0) \} \, , \\[2mm]
U_{(a,b)} & \text{constant and positive} & (a, b) = \{ (2, 2), (1, 0) \} \, , \\[2mm]
U_{(a,b)} & \text{can be stabilised with a positive value} & (a, b) = \{ (4, 3), (3, 3), (1, 1) \} \, 
\end{array}
\end{align}
We could check whether there are critical points at $\varphi_2 = 0$. We find that 
\be
\partial_{\varphi_2} U_{(p,q)} |_{\varphi_2 = 0} = 0 \qquad \Longleftrightarrow \qquad p = q \, .
\ee
There is also a critical point for $U_{(p,q)}$ when $p=4, q=3$, but it is not in the origin\footnote{By the origin of moduli space we always mean the point $\vec{\phi} = \vec{0}$.} ($\varphi_2 = 0$). Let us focus on the $p=q$ case as an illustration. The value of $U_{(p,q)}(\varphi_2 = 0)$ is
\be
U_{(p, q)}(\varphi_2 = 0) = (p+q) - \tfrac{1}{2} \, (p-q)^2\,.
\ee
When $p=q, U_{(p, q)}$ is always equal to $2 \, p$ and hence positive. This means that the full potential becomes 
\begin{equation}
V_{\textsc{old}}(\varphi_1,0) = 2 \, p \, \e^{2 \, r \, \varphi_1} \, .
\end{equation}
When we write $\varphi_1$ in terms of a canonically normalized scalar 
\be 
\varphi_1 = a_1 \, \phi_1 \, ,  
\ee
with $a_1$ as some constant to be determined below, in order to satisfy the requirements, we must have
\be 
(r \, a_1)^2 < \frac{3}{4} \, .
\ee
The power-law cosmological solution now reads
\be
\d s^2 = -\d t^2 + a(t)^2\d\vec{x}^2\,,\qquad a(t)= t^{\frac{1}{(2 \, r \, a_{1})^2}} \,.
\ee
Let us look at the scalar kinetic term in the 4D action and fix the value of $a_1$. It is given by
\be
S=\int_4 \sqrt{-g} \, \Bigl( \tfrac{1}{4} \, \Tr \big\{ \star\d \mathcal{M} \wedge \d \mathcal{M}^{-1} \big\} \Bigr)\,.
\ee
For the kinetic term we find 
\be
\tfrac{1}{4}\Tr \big\{ \star\d \mathcal{M} \wedge \d \mathcal{M}^{-1} \big\} = -2 \, r \, (p+q) \, a_1^2 \, (\partial\varphi_{1})^2 \quad \Longrightarrow \quad a_1^2 = \frac{1}{4 \, r \, (p+q)}\,,
\ee
when only $\varphi_1$ is non-constant. Hence, we find the following scaling cosmologies
\be
a(t) = t^{\frac{p+q}{r}}\,,
\ee
which means we have cosmic acceleration if $p+q>r$. This means that we have only one possibility with $r \neq 0$, namely $p = q = 3$. This is the scaling solution found in the ${\rm{CSO}}(3,3,2)$ gauged maximal supergravity in \cite{Townsend:2001ea}.

\subsection{The power-law solutions for the new theories}
Let us now consider the new $\omega$-deformed maximal supergravities. In the construction of the scalar potential, we follow the appendix of \cite{Borghese:2012zs}. We can express it in terms of a complex superpotential
\be
W = \Tr \big\{ \cos (\omega) \, \eta \, \mathcal{M} -i \, \sin (\omega) \, \eta' \, \mathcal{M}^{-1} \big\} \, ,
\ee
as follows
\be
V = - \, \tfrac{3}{8} \, |W|^2 + \tfrac{1}{2} \, g^{ij} \, \partial_i W \, \partial_j \overline{W}\,.
\ee
The moduli metric $g_{ij}$ is defined in such a way that
\be
\tfrac{1}{4} \, \Tr \big\{ \partial \mathcal{M} \, \partial \mathcal{M}^{-1} \big\} =  \tfrac{1}{2} \sum_{ij} g_{ij} \, \partial \varphi_i \, \partial\varphi_j \, .
\ee
For the $\omega$-deformed ${\rm{CSO}}(p,q,r)$ gaugings, we can choose same $\eta$ as in the previous section (\ref{eta def}), while $\eta'$ must be chosen in order to satisfy the constraint 
\be \label{constraint}
\eta \, \eta' = \mathbb{0}_8 \, . 
\ee
This is the only surviving condition among the so-called quadratic constraints, which are necessary to ensure consistency of the gauging procedure. We will choose
\be
\eta'=\begin{pmatrix} \mathbb{0}_{p+q} & \\
 & \mathbb{D}_r
\end{pmatrix},
\ee
with  $r=s+t+u$ and $\mathbb{D}_r$ a diagonal matrix, with $1, -1, 0$  as entries and is given by
\be
\mathbb{D}_r=\begin{pmatrix}
\mathbb{1}_s & & \\
 & - \mathbb{1}_t &  \\
& & \mathbb{0}_u
\end{pmatrix}
\ee
With these choices the quadratic constraint (\ref{constraint}) is trivially satisfied. If we write down the scalar potential for the $\omega$ deformed supergravities explicitly, we find the following interesting relation
\be
V = \cos^2 (\omega) \, V_{\textsc{old}}[\eta, \mathcal{M}] + \sin^2 (\omega) \, V_{\textsc{old}}[\eta', \mathcal{M}^{-1}] \, .
\ee
This means that the theory is really a superposition, with weights $\cos^2\omega, \sin^2\omega$ of two different gaugings: ${\rm{CSO}}(p,q,r)$ and ${\rm{CSO}}(s, t, u+p+q)$. The second gauging uses the `inverse' (or dual) coset representation for the scalars, compared to the first gauging, since the second piece of the potential is a function of $\mathcal{M}^{-1} = \mathcal{L}^{-T} \, \mathcal{L}^{-1}$, where $\mathcal{L}$ is the usual coset representative.

\subsubsection{Analysis of the scalar potential}
{\bf Simple case: $s=r$}\\[2mm]
Consider first the case $\mathbb{D}_r=\mathbb{1}_r$. In this case we expect the previous truncation to $\varphi_1, \varphi_2$ to be consistent. In other words, we are analysing the singlets under $\SO(p) \times \SO(q) \times \SO(r),$ but the scalar potential is modified by the $\omega$ parameter in the following way: 
\be
V(\varphi_{1}, \varphi_{2}) = \cos^2 (\omega) \, U_{(p,q)}(\varphi_{2}) \, \e^{2 \, r \, \varphi_{1}} + \sin^2 (\omega) \, \tfrac{(2-r) \, r}{2} \, \e^{2\, (8-r) \, \varphi_{1}} \, ,
\ee
with $V_{\text{old}}$ as given in (\ref{standard1}, \ref{standard2}). We expect the exponential runaway behavior of $\varphi_1$ to be lifted, unless $r=4$. There could even be the possibility of stabilising the field $\varphi_{1}$ obtaining an actual vacuum and not a scaling solution. Let us briefly go through all the possibilities:
\begin{itemize}
\item  When $r=1$, we take $(p,q) = (4,3)$. Then, for a given finite value $\varphi_{2} = \varphi_{2}^{\star},$ we can stabilise $U_{(4,3)}(\varphi_{2})$ say to $\Lambda = U_{(4,3)}(\varphi_{2}^{\star}) > 0$. It remains to see whether it is possible to stabilise the scalar $\varphi_{1}$. This means that we need to find the critical points of
\be
V(\varphi_{1}) = \Lambda \, \cos^2 (\omega) \, \e^{2 \, \varphi_{1}} + \tfrac{1}{2} \, \sin^2 (\omega) \, \e^{14 \, \varphi_{1}} \, .
\ee
Unfortunately, this function does not have a critical point for finite $\varphi_{1}$.
\item When $r=2$, the extra term cancels and we still have the power-law solution $a (t)\sim t^3$. However now the value of the scalar potential is scaled by a factor $\cos^{2} (\omega)$.
\item When $r=3$, $U_{(5,0)}$ is constant and negative. But now the extra term containing $\sin^2\omega$ also becomes negative, and there is no dS solution at finite $\varphi_{1}$.
\item When $r=4$, the extra term does not vanish, but has the same $\varphi_{1}$ dependence. So, we have new power-law solutions with $a(t)\sim t$. Something special happens for $p=2$ and $q=2$. In this case, $U_{(2,2)}(\varphi_{2})$ is constant and positive. The first and second term in the potential are constants and have the same value 
but with opposite signs. The potential looks like
\be
V(\varphi_{1}) = \e^{8 \, \varphi_{1}} \left [ 4 \, \cos^2 (\omega)  -4 \, \sin^2 (\omega) \right ] = 4 \, \e^{8 \, \varphi_{1}} \cos (2 \, \omega) \, ,
\ee
This solution is similar to the one we found in the old theory. The only difference is the modulation by a cosine factor which allows us to tune the scalar potential to whatever positive value $-4 < \Lambda_0 < 4$.
\item From similar considerations, when $r>4$, there are no new solutions.
\end{itemize}
{\bf General case}\\[2mm]
\noindent Now let us consider the general case where
\be
\eta' =\begin{pmatrix}
 \mathbb{0}_{p+q} & & & \\
  & \mathbb{1}_s & & \\
& & -\mathbb{1}_t & \\
& & & \mathbb{0}_u  
\end{pmatrix},
\ee
where, $p+q+r=8$ and $r= s+t+ u$ as usual. It is straightforward to see that we will generically have a \emph{four}-scalar system. We will choose the following parametrisation of the scalar metric $\mathcal{M}:$
\begin{align} \mathcal{M} = \e^{r \, \varphi_{1}} \begin{pmatrix}
 \e^{q \, \varphi_{2}} \, \mathbb{1}_p & & \\
 & \e^{-p \, \varphi_{2}} \, \mathbb{1}_q & \\
 & & \e^{-8 \, \varphi_{1}} \, \begin{pmatrix} \e^{t \, \varphi_{4} - u \, \varphi_{3}} \, \mathbb{1}_s & & \\
& \e^{-s \, \varphi_{4} - u \, \varphi_{3}} \, \mathbb{1}_t & \\
& & \e^{(t+s) \, \varphi_{3}} \, \mathbb{1}_u \end{pmatrix}
\end{pmatrix}.
\end{align}
The nice feature about this parametrisation is that there are no cross terms for the scalars in the kinetic terms. This is necessary for understanding the existence of the power-law solutions. This set of fields corresponds to the set of $\SO(p) \times \SO(q) \times \SO(s) \times \SO(t) \times \SO(u)$ singlets. In terms of canonically normalized scalars, we have
\be
 \varphi_{1} =  a_1 \, \phi_1\quad , \quad  \varphi_{2} = a_2 \, \phi_2 \quad , \quad \varphi_{3} = a_{3} \, \phi_3 \quad ,\quad \varphi_{4} = a_4 \, \phi_4 \, ,
\ee
with
\begin{align}
a_1^2= \frac{1}{4 \, r \, (p+q)} \quad , \quad a_2^2 = \frac{2}{p \, q \, (p+q)} \quad , \quad a_3^2 = \frac{2}{u \, r \, (r-u)} \quad , \quad a_4^2 = \frac{2}{s \, t \, (s+t)} \, .
\end{align}
In full generality, the scalar potential should be seen as a function of two exponentials 
\be
V = \cos^2 (\omega) \, U_{(p,q)}(\varphi_{2}) \, \e^{2 \, r \, \varphi_{1}} + \sin^2 (\omega) \, U_{(s,t)}(-\varphi_{4})\, \e^{2\, (8-r) \, \varphi_{1} + 2 \, u \, \varphi_{3}} \, , 
\ee
where $U_{(p, q)}$ and $U_{(s,t)}$ have been defined in equation (\ref{standard2}).

\subsubsection{Candidates for acceleration: Positive potential with $P>1$}
If both $U_{(p,q)}(\varphi_{2})$ and $U_{(s,t)}(-\varphi_{4})$ can be extremised at positive values, the potential can be written in terms of canonical scalars as
\be
V(\phi_1, \phi_3) = \Lambda_1 \, \cos^2 (\omega) \, \e^{2 \, r \, a_1 \phi_1} + \Lambda_2 \, \sin^2 (\omega) \, \e^{2 \, (8-r) \, a_1 \phi_1 + 2 \, u \, a_3 \phi_3} \, ,
\ee
where $\Lambda_1, \Lambda_2 >0$. In \cite{Hartong:2006rt}, the possibility of realising a scaling cosmology starting from a similar scalar potential was analysed . In principle, it is possible to find a scaling solution with the following power-law
\be
a \sim t^P\,,
\ee
where $P$ is the sum of all the matrix elements of $A^{-1}$, where $A$ is defined as
\be
A=\begin{pmatrix} 
4 \, r^2 \, a_1^2   & 1 \\
1 &  4 \, (8-r)^2 \, a_1^2 +4 \, u^2 \, a_3^2
\end{pmatrix}.
\ee
Below we present a list of possible inequivalent combinations of $(p,q)$ and $(s, t)$ that might generate new scaling solutions.
\begin{table}[ht] 
\centering \begin{tabular}{c c c} 
\hline\hline
Case no.&$(p,q) + (s,t)$& $\phantom{\dfrac{1}{2}}P$ \\[1mm] 
\hline \\[-8mm] & & \\
1&$(3, 3) + (1, 0)$&$4$ \\[1mm]
2&$(2, 2) + (1, 1)$&$1$ \\[1mm]
3&$(2, 2) + (1, 0)$&$1$ \\[1mm] 
4&$(1, 1) + (1, 1)$&$\nicefrac{1}{2}$ \\[1mm]
5&$(1, 1) + (1, 0)$&$\nicefrac{2}{5}$ \\[1mm]
6&$(1, 0) + (1, 0)$&$\nicefrac{1}{4}$ \\[1mm] 
\hline \hline \end{tabular} 
\caption{Candidates for accelerated scaling solutions with related power $P$} 
\end{table}

Despite the fact that the first three cases might represent scaling solutions, they are ruled out by a consistency condition in \cite {Hartong:2006rt}. For a detailed analysis, we refer to Appendix A. The last three cases give new consistent power-law solutions, but they are not accelerating since $P<1$.

\subsubsection{Candidates for ekpyrosis: Negative potential with $P<\frac{1}{3}$}  

Ekpyrosis would require negative exponential with power-law $P<\frac{1}{3}$. 
The possible inequivalent combinations $(p,q)$ and $(s,t)$ that give new scaling solutions are given in Table 6. In all these cases, $P > \frac{1}{3}$. So, these solutions are not viable ekpyrotic ones.
\begin{table}[h!] 
\centering \begin{tabular}{c c c} 
\hline\hline
Case no.&$(p,q) + (s,t)$& $\phantom{\dfrac{1}{2}}P$ \\[1mm] 
\hline \\[-8mm] & & \\
1&$(4, 0) + (3, 0)$&$1$ \\[1mm]
2&$(3, 0) + (3, 0)$&$\nicefrac{3}{4}$ \\[1mm] 
\hline \hline \end{tabular} 
\caption{Candidates for ekpyrotic solutions with related power $P$} 
\end{table}

Some of the mixed combinations of $U$ might also be possible candidates for Ekpyrosis. We explored all those combinations and did not find any cases with $P<\frac{1}{3}$. A detailed summary of the possible combinations can be found in Appendix A.

\section{Discussion}

In this paper, we have investigated the possibility of obtaining cosmological power-law solutions for maximal and minimal supergravity theories. Our goal was to search for both accelerating and ekpyrotic scaling solutions. 

We have first considered the {\it STU} model for minimal supergravity. We have searched for accelerating power-law solutions. So far it has not been possible to get a stable dS solution for this type of model without adding non geometric fluxes. However, adding these fluxes might make the supergravity theory untrustworthy; so, scaling solutions would be the next best choice. We have also studied ekpyrotic solutions for this model since these are the most popular altenatives for cosmological inflation and the mathematical analysis is similar to the search for  accelerating solutions. 

In this {\it STU} model, the full scalar potential has eleven expoential terms with six scalars (three dilatons and three axions). Besides the NSNS and RR fluxes, the theory also has metric fluxes generating the scalar potential.
We want to extract an overall exponential factor for at least one field after a field rotation, if necessary. We have studied all possible combinations of the largest common sets of $\vec{\alpha}_i$ (\ref{app:stu}) that are mutually affine. We have identified 17 possible sets of mutually affine vectors. The $P$ value for each of these sets corresponds to accelerating, ekpyrotic or generic scaling solutions. By choosing the appropriate values for the fluxes, we can keep a specific set of mutually affine vectors. Out of these 17 possible cases, there are only three cases which represent possible candidates for accelerating scaling solutions with $P>1$. On the other hand, every single set of mutually affine vectors represents a candidate for ekpyrotic solutions because of the fact that, by switching off some fluxes, we can lower the $P$ value.
We did not find any accelerating solution. However, we found three fully stabilized ekpyrotic solutions with $P=\nicefrac{3}{10},\nicefrac{5}{22}$ and $\nicefrac{13}{54}$. 

Very recently, it has been proposed that there exists a new one-parameter family of SO(8) gauged $\mathcal{N} = 8$ supergravity theories \cite{Dall'Agata:2012bb}. A similar one-parameter generalisation can be considered for ${\rm{CSO}}(p,q,r)$ gauged maximal supergravities. For such theories, the scalar potential for a truncated subset of fields parametrising the coset $\SL(8) / \SO(8)$ has been constructed in \cite{Borghese:2012zs}. The scalar potential depends on the new parameter, which we denote with $\omega$. In this paper, we have considered a further truncation of the ${\rm{CSO}}(p,q,r)$ theories to singlets under $\SO(p) \times \SO(q) \times \SO(s) \times \SO(t) \times \SO(u)$. The truncation contains at most four scalar fields.

We have studied the possibility of realising accelerating scaling solutions or ekpyrotic solutions within these new theories. We have first considered a special case and found a new generic scaling solution with $P=1$. Then, we have explored the most general case. Among six candidate accelerating solutions, only one could have actually been an accelerated scaling solution having $P>1$. Looking into the details of the case, we have found that the function multiplying the running scalar  cannot be stabilized. For two other cases, we have found new generic scaling solutions with $\frac{1}{3}<P<1$. We have followed a similar approach for the ekpyrotic solutions, and identified two cases that would have given us possible solutions. But neither of the cases has a $P<\frac{1}{3}$; so, they do not qualify for further tests. We have analysed all other possible combinations of $\{p,q,s,t,u\}$ finding no viable scaling or ekpyrotic solutions. Our analysis is exhaustive when restricted to singlets under $\SO(p) \times \SO(q) \times \SO(s) \times \SO(t) \times \SO(u)$. 

Despite the amount of considered cases, we have not been able to find any new accelerating scaling solution. This is probably due to the large symmetry group preserved by scalars in our truncation. A similar consideration could be done for dS critical points in the new $\SO(4,4)$ gauged supergravities \cite{Dall'Agata:2012sx}. As already mentioned, in \cite{Dall'Agata:2012sx} it has been shown that tachyons can be made arbitrarily light by tuning the $\omega$ parameter. However, this is not true for the critical point at the origin of moduli space preserving $\SO(4) \times \SO(4),$ but only for critical points away from the origin preserving a smaller amount of symmetry, namely $\SO(3) \times \SO(3)$. Perhaps, reducing the dimension of the preserved symmetry group will lead to new solutions.

Another open issue is the higher dimensional origin of the $\omega$ deformed theories. One of the original motivations for considering scaling cosmologies in maximal supergravity theories was the clear link between the latter and M-theory. In some sense, having found a scaling solution within the old maximal supergravity meant having embedded a scaling universe in a higher dimensional theory. This link is less clear for the new supergravities. Very recently, a first attempt to pinpoint the higher dimensional origin of the $\SO(8)$, $\omega$ deformed, gauged theory has been put forward \cite{deWit:2013ija}, but the issue remains unsolved.

\section{Acknowledgement} 
We would like to thank Thomas Van Riet and Diederik Roest for useful explanations and discussions on this topic. We also thank Pawel Caputa, Giuseppe Dibitetto, Vishnu Jejjala, Ilies Messamah for useful discussions. JB is supported by the Swedish Research Council (VR), and the G\"oran Gustafsson Foundation. The work of SSH is supported by the South African Research Chairs Initiative of the Department of Science and Technology and National Research Foundation. AB gratefully acknowledges support by a VIDI grant from NWO.
\appendix

\section{Appendix}

\subsection{$\vec \alpha$-vectors from the {\it STU} potential} \label{app:stu} \ \
From the full scalar potential we get the following $\vec \alpha$ vectors \cite{Collinucci:2004iw} :
\begin{align}
& \vec \alpha_1=(0, \sqrt{6}, 0), \ \vec \alpha_2=(\sqrt{2}, 2 \sqrt{\nicefrac{2}{3}}, 0),\  \vec\alpha_3=(\sqrt{2}, \sqrt{\nicefrac{2}{3}}, \sqrt{\nicefrac{2}{3}}),\ \ \vec\alpha_4=(0, 2\sqrt{\nicefrac{2}{3}}, \sqrt{\nicefrac{2}{3}})\,,\nonumber\\
&\vec \alpha_5=(-\sqrt{2}, \sqrt{6}, \sqrt{\nicefrac{2}{3}}), \ \vec\alpha_6=(\sqrt{2}, \sqrt{6}, -\sqrt{6})\ \ \vec \alpha_7=(\sqrt{2}, \sqrt{6}, -\sqrt{\nicefrac{2}{3}}), \ \vec \alpha_8=(-\sqrt{2}, \sqrt{6}, \sqrt{6}) \,,\nonumber\\
& \vec \alpha_9=(\sqrt{2}, \sqrt{\nicefrac{2}{3}}, \sqrt{6}),\ \  \vec \alpha_{10}=(\sqrt{2},   \sqrt{6}, \sqrt{\nicefrac{2}{3}}), \ \vec \alpha_{11}=(\sqrt{2}, \sqrt{6}, \sqrt{6})\,.
\end{align}
\subsection{Stability analysis for possible solutions in new Maximal supergravity}\ \

\noindent {\underline{\bf $(3, 3) + (1, 0)$ Case:}} \\[2mm]
Here $p=3, q=3$ and $s=1, t=0$. Hence $r=2$ and $u=1$. In terms of normalised scalar fields we get 
\be
V =\cos^2 (\omega) \, U_{(3,3)}(\phi_{2}) \, \e^{\frac{1}{\sqrt{3}} \, \phi_3} + \sin^2 (\omega) \, U_{(1,0)}(-\phi_{4}) \, \e^{\sqrt{3} \, \phi_{1} + 2 \, \phi_{3}}
\ee
and 
\begin{align*}
U_{3, 3}(\phi_{2}) & = - \tfrac{3}{2} \, \e^{\frac{2 \, \phi_{2}}{\sqrt{3}}} - \tfrac{3}{2} \, \e^{- \frac{2 \, \phi_{2}}{\sqrt{3}}} + 9 \, , \\
U_{1, 0}(-\phi_{4}) & = \tfrac{1}{2} = \Lambda_2 \, .
\end{align*}
We would like to write the scalar potential as $V=\e^{\alpha \, \psi_1} \, U(\phi_2, \psi_{3})$. To get a common factor we transform the fields $\{\phi_{1}, \phi_{3}\}$ by an $\SO(2)$ matrix in order to preserve the normalisation. We obtain
\begin{align*}
V(\psi_{1}, \phi_{2}, \psi_{3}) & = \e^{\frac{1}{2} \, \psi_{1}} \, U(\phi_{2}, \psi_{3}) \\
& =  \left[  \cos^2 (\omega) \, \e^{\frac{1}{2 \, \sqrt{3}} \, \psi_{3}} \big( - \tfrac{3}{2} \, \e^{\frac{2 \, \phi_{2}}{\sqrt{3}}} - \tfrac{3}{2} \, \e^{- \frac{2 \, \phi_{2}}{\sqrt{3}}} + 9 \big) + \tfrac{1}{2} \, \sin^2 (\omega) \, \e^{\frac{3 \, \sqrt{3}}{2} \, \psi_{3}} \right] \, .
\end{align*}
From section 3.1 we know that $U_{(3,3)}(\phi_{2})$ can be stabilised. In particular we have a stationary point for $\phi_{2}^{\star} = 0$ and $U_{(3,3)}(0) = \Lambda_1 = 6$. Thus we are left with
\begin{align*}
V(\psi_{1}, \psi_{3}) & = \e^{\frac{1}{2} \, \psi_{1}} \, U(\psi_{3}) \\
& = \e^{\frac{1}{2} \, \psi_{1}} \, \left[ 6 \, \cos^2 (\omega) \, \e^{\frac{1}{2 \, \sqrt{3}} \, \psi_{3}} + \tfrac{1}{2} \, \sin^2 (\omega) \, \e^{\frac{3 \, \sqrt{3}}{2} \, \psi_{3}} \right] \, .
\end{align*}
For any value of $\omega$ the function $U$ is unstable. The scaling solution is thus inconsistent. \\

\noindent {\underline{\bf $(2, 2) + (1, 1)$ Case: }}\\[2mm]
Here, $p=2, q=2$ and $s=1, t=1$ then $r=4$ and $u=2$. The scalar potential is
\be
V = \cos^2(\omega) \, U_{(2, 2)}(\phi_{2}) \, \e^{\phi_1} + \sin^2(\omega) \, U_{(1,1)}(-\phi_{4}) \, \e^{\phi_1+ \sqrt{2} \, \phi_3} \, ,
\ee
with 
\begin{align*}
U_{(2, 2)}(\phi_{2}) & = 4 = \Lambda_1 \, , \\
U_{(1, 1)}(-\phi_{4}) & = 2 \, \cosh^2 (\phi_{4}) \, .
\end{align*}
We can again stabilise $U_{(1,1)}$ to $\Lambda_2 = U_{(1,1)}(0) = 2$ thus obtaining
\begin{align*}
V(\phi_{1}, \phi_{3}) & = \Lambda_1 \, \cos^{2} (\omega) \, \e^{\phi_1} + \Lambda_{2} \, \sin^2(\omega) \,  \e^{\phi_1+ \sqrt{2} \, \phi_3} \, , \\
& = 2 \, \e^{\phi_{1}} \, \left[ 2 \, \cos^{2} (\omega) + \sin^2(\omega) \, \e^{\sqrt{2} \, \phi_3} \right] \, .
\end{align*}
Unfortunately it is impossible to stabilise $\phi_{3}$ for $0 < \omega < \pi$.\\

\noindent {\underline{\bf $(2, 2) + (1, 0) $ Case: }}\\[2mm]
Here, $p=2, q=2$ and $s=1, t=0$ then $r=4$ and $u=3$.
The scalar potential will look like:
\be
V = \cos^2 (\omega) \, U_{(2, 2)}(\phi_{2}) \, \e^{\phi_1} + \sin^2 (\omega) \, U_{(1,0)}(-\phi_{4}) \, \e^{\phi_1 + \sqrt{6}\, \phi_3} \, ,
\ee
with
\begin{align*}
U_{(2,2)}(\phi_{2}) & = 4 = \Lambda_1 \, , \\
U_{(1,0)}(-\phi_{4}) & = \tfrac{1}{2} = \Lambda_2 \, .
\end{align*}
In this case
\begin{align*}
V = \tfrac{1}{2} \, \e^{\phi_{1}} \, \left[ 8 \, \cos^{2} (\omega) + \sin^{2} \omega \, \e^{\sqrt{6} \, \phi_{3}} \right] \, .
\end{align*}
It is impossible to stabilise $\phi_{3}$ for $0 < \omega < \pi$.\\

\noindent {\underline{\bf $(1, 1) + (1, 1)$ Case:}}\\[2mm]
Here, $p=1,q=1$ and $s=1,t=1$ meaning $r=6$ and $u=4$. The scalar potential is
\be
V = \cos^2 (\omega) \, U_{(1,1)}(\phi_{2}) \, \e^{\sqrt{3} \, \phi_1} + \sin^2(\omega) \, U_{(1,1)}( -\phi_{4}) \, \e^{\frac{1}{\sqrt{3}} \, \phi_1 + \frac{2 \sqrt{2}}{\sqrt{3}} \, \phi_3} \, .
\ee
with
\begin{align*}
U_{(1,1)}(\phi) = U_{(1,1)}(- \phi) = 2 \, \cosh^{2} (\phi) \, .
\end{align*}
We can stabilise both $U_{(1,1)}(\phi_{2})$ and $U_{(1,1)}(- \phi_{4})$ to $\Lambda_{1} = \Lambda_{2} = 2$. After an $\SO(2)$ transformation on $\{\phi_{1} ,\phi_{3}\}$ we are left with
\begin{align*}
V = \e^{\sqrt{2} \, \psi_1} \, U(\psi_{3}) = 2 \, \e^{\sqrt{2} \, \psi_1} \, \big[ \cos^{2} (\omega) \, \e^{-\psi_{3}} + \sin^{2} (\omega) \, \e^{\psi_{3}} \big] \, .
\end{align*}
For any non trivial value of $\omega$ the function $U(\psi_{3})$ could be stabilised giving a consistent scaling solution. The power-law exponent is $P = 1$ and hence the solution is not accelerating. \\

\noindent {\underline{\bf $(1, 1) + (1, 0) $  Case: }}\\[2mm]
Here, $p=1,q=1$ and $s=1,t=0$ then $r=6$ and $u=5$. The scalar potential is
\be
V =\cos^2 (\omega) \, U_{(1, 1)}(\phi_{2}) \, \e^{\sqrt{3} \, \phi_{1}} + \sin^2 (\omega) \, U_{(1, 0)}( -\phi_{4}) \, \e^{\frac{1}{\sqrt{3}} \, \phi_{1} + \frac{2 \sqrt{5}}{\sqrt{3}} \, \phi_{3}} \, .
\ee
with
\begin{align*}
U_{(1, 1)}(\phi_{2}) & = 2 \, \cosh^{2}(\phi_{2}) \, , \\
U_{(1, 0)}(-\phi_{4}) & = \tfrac{1}{2} = \Lambda_{2} \, .
\end{align*}
After having stabilised $U_{(1,1)}(\phi_{2})$ and an $\SO(2)$ rotation on $\{ \phi_{1} , \phi_{3} \}$ we find
\begin{align*}
V = \e^{\sqrt{\frac{5}{2}} \, \psi_{1}} \, U(\psi_{3}) = \tfrac{1}{2} \, \e^{\sqrt{\frac{5}{2}} \, \psi_{1}} \left[ 4 \, \cos^{2} (\omega) \, \e^{- \frac{1}{\sqrt{2}} \, \psi_{3}} + \sin^{2} (\omega) \, \e^{2 \, \psi_{3}} \right] \, .
\end{align*}
The function $U$ could again be stabilised for any non trivial value of $\omega$ giving a scaling solution. In this case $P = \nicefrac{2}{5}$ meaning the solution is not accelerating. \\

\noindent {\underline{\bf $(1, 0) + (1, 0) $  Case: }}\\[2mm]
Here, $p=1,q=0$ and $s=1,t=0$ then $r=7$ and $u=6$. The scalar potential is
\be
V =\cos^2 (\omega) \, U_{(1, 0)}(\phi_{2}) \, \e^{\sqrt{7} \, \phi_{1}} + \sin^2 (\omega) \, U_{(1, 0)}( -\phi_{4}) \, \e^{\frac{1}{\sqrt{7}} \, \phi_{1} + \frac{4 \sqrt{3}}{\sqrt{7}} \, \phi_{3}} \, .
\ee
with $U_{(1,1)}(\phi_{2}) = \tfrac{1}{2} = \Lambda_1$ and $U_{(1,1)}(-\phi_{4}) = \tfrac{1}{2} = \Lambda_2$. After an $\SO(2)$ rotation on $\{ \phi_{1} , \phi_{3} \}$ we find
\begin{align*}
V = \tfrac{1}{2} \, \e^{2 \, \psi_{1}} \, U(\psi_{3}) = \tfrac{1}{2} \, \e^{2 \, \psi_{1}} \left[ \cos^{2} (\omega) \, \e^{- \sqrt{3} \, \psi_{3}} + \sin^{2} (\omega) \, \e^{\sqrt{3} \, \psi_{3}} \right] \, .
\end{align*}
The function $U$ could again be stabilised to a positive value for any non trivial value of $\omega$ giving a scaling solution. In this case $P = \nicefrac{1}{4}$ meaning the solution is not accelerating. Note that, being $P = \nicefrac{1}{4} < \nicefrac{1}{3}$ if the value of $U$ had been negative we could have used this solution to realise ekpyrosis. Unfortunately this is not the case. \\

\noindent {\bf Mixed combinations of $U$ for Ekpyrosis: }\\[2mm]
We are looking at the nontrivial combinations of $U$'s with $\Lambda_1 = U_{(p,q)}(\varphi_{2}^{\star}) < 0$ and $\Lambda_2 = U_{(s,t)}(\varphi_{4}^{\star}) > 0$. The possibilities are summarised in Table 9.
\begin{table}[ht] 
\centering \begin{tabular}{c c c c} 
\hline\hline
Case no.&$(p,q) + (s,t)$& $\phantom{\dfrac{1}{2}}P$ & Comments \\[1mm] 
\hline \\[-8mm] & & & \\
1&$(6, 0) + (1, 0)$&$4$ & no ekpyrosis \\[1mm]
2&$(5, 0) + (1, 0)$&$\nicefrac{7}{4}$ & no ekpyrosis \\[1mm]
3&$(5, 0) + (1, 1)$&$2$ & no ekpyrosis \\[1mm] 
4&$(4, 0) + (1, 0)$&$1$ & no ekpyrosis \\[1mm]
5&$(4, 0) + (1, 1)$&$1$ & no ekpyrosis \\[1mm]
6&$(3, 0) + (1, 0)$&$\nicefrac{5}{8}$ & no ekpyrosis \\[1mm]
7&$(3, 0) + (1, 1)$&$\nicefrac{2}{3}$ & no ekpyrosis \\[1mm]
8&$(3, 0) + (2, 2)$&$1$ & no ekpyrosis \\[1mm] 
\hline \hline \end{tabular} 
\caption{Mixed cases} 
\end{table}
Please note that, at first sight, in the list the first three cases are excellent candidates for accelerating solutions. After a closer look we find that they are stabilized only at negative potential. By the same argument we ruled out case $(1, 0)+(1, 0)$ for ekpyrosis, since it has a positive $U.$ \\

\noindent {\bf Special cases}\\[2mm]
\begin{table}[ht] 
\centering \begin{tabular}{c c c} 
\hline\hline
Case no.&$(p,q) + (s,t)$& $\phantom{\dfrac{1}{2}}P$ \\[1mm] 
\hline \\[-8mm] & & \\
1&$(2, 2) + (2, 0)$&$1$ \\[1mm]
2&$(1, 1) + (2, 0)$&$\nicefrac{1}{3}$ \\[1mm] 
3&$(1, 0) + (2, 0)$&$\nicefrac{1}{7}$ \\[1mm]
\hline \hline \end{tabular} 
\caption{Candidates for accelerating scaling solutions with related power $P$} 
\end{table}
\begin{table}[ht] 
\centering \begin{tabular}{c c c} 
\hline\hline
Case no.&$(p,q) + (s,t)$& $\phantom{\dfrac{1}{2}}P$ \\[1mm] 
\hline \\[-8mm] & & \\
1&$(5, 0) + (2, 0)$&$\nicefrac{5}{3}$ \\[1mm]
2&$(4, 0) + (2, 0)$&$1$ \\[1mm] 
3&$(3, 0) + (2, 0)$&$\nicefrac{3}{5}$ \\[1mm]
\hline \hline \end{tabular} 
\caption{Candidates for ekpyrotic solutions with related power $P$} 
\end{table}

We could consider the special case in which either $U_{(p,q)}(\varphi_{2})$ or $U_{(s,t)} (-\varphi_{4})$ is zero. This corresponds to the choice $(a,b) = (2,0)$. In this case the scalar potential is similar to that of the old supergravity theories but modulated by a $\cos^{2}(\omega)$ or $\sin^{2}(\omega)$ factor. In Tables 8 and 9 we list all relevant cases. \\

\providecommand{\href}[2]{#2}\begingroup\raggedright\endgroup

\bibliographystyle{utphysmodb}

\end{document}